# Observation of Kondo condensation in a degenerately doped silicon metal

H. Im[1,2], D. U. Lee[3], Y. Jo[1], J. Kim[1], Y. Chong[4], W. Song[5], H. Kim[1], E. K. Kim[3], S.-J. Sin[3], S. Moon[3], J. R. Prance[6], Yu. A. Pashkin[6,†] and J. S. Tsai[2,7]

[1] *Division of Physics and Semiconductor Science, Dongguk University, Seoul 04620, Korea*
[2] *The Institute of Physical and Chemical Research (RIKEN), 34 Miyukigaoka Tsukuba, Ibaraki 305-8501, Japan*
[3] *Department of Physics, Hanyang University, Seoul 04763, Korea*
[4] *SKKU Adavanced Institute of Nanotechnology (SAINT), SungKyunKwan University, Suwon 16419, Korea*
[5] *Korea Research Institute of Standards and Science, Daejeon 34113, Korea*
[6] *Department of Physics, Lancaster University, Lancaster, LA1 4YB, UK*
[7] *Department of Physics, Tokyo University of Science, Kagurazaka, Tokyo 162-8601, Japan*

**When a magnetic moment is embedded in a metal, it captures itinerant electrons to form the Kondo cloud[1,2], which can spread out over a few micrometres[3,4]. For a metal with dense magnetic impurities such that Kondo clouds overlap with each other, correlated ground states are formed. When the impurities form a regular lattice, the result is a heavy fermion or anti-ferromagnetic order depending on the dominant interaction[5,6]. Even in the case of random impurities, overlapping Kondo clouds are expected to form a coherent ground state. Here, we examine this issue by performing electrical transport and high-precision tunnelling density-of-states (DOS) spectroscopy measurements in a highly P-doped crystalline silicon metal where disorder-induced localized magnetic moments exist[7]. We detect the Kondo effect in the resistivity of the Si metal below 2 K and an exotic pseudogap in the DOS with gap edge peaks at a Fermi energy below 100 mK. The DOS gap and peaks are tuned by applying an external magnetic field and transformed into a metallic Altshuler-Aronov gap[8] in the paramagnetic disordered Fermi liquid (DFL) phase. We interpret this phenomenon as the Kondo condensation, the formation of a correlated ground state of overlapping Kondo clouds, and its transition to a DFL. The boundary between the Kondo condensation and DFL phases is identified by analysing distinct DOS spectra in the magnetic field-temperature plane. A detailed theoretical analysis using a holographic method[9,10,11] reproduces the unusual DOS spectra,**

**supporting our scenario. Our work demonstrates the observation of the magnetic version of Bardeen-Cooper-Shrieffer (BCS) pair condensation and will be useful for understanding complex Kondo systems.**

The interplay of electron-electron interactions, disorder, and spin correlation in solids is the origin of the existence of many competing ground states and phase transitions between them[6,12,13], which are typically observed in stongly correlated complex materials ranging from heavy fermion compounds[14] to high-temperature superconductors[15,16]. An intriguing question in complex interacting systems is how the interactions of microscopic particles lead to macroscopic phenomena such as superconductivity, charge and spin density waves, and heavy fermions. The Kondo effect very often plays a central role in understanding the correlated ground states of electron systems (see Fig. 1). Doping is a versatile tool to address this question because of its ability to control the interaction strength and the way in which particles interact with each other with external perturbations such as magnetic fields and pressure[17,18].

Here, we report the observation of the Kondo interaction and an exotic Bardeen-Cooper-Shrieffer (BCS)-type pseudogap in ultra-highly P-doped degenerate silicon (Si:P) with doping concentrations of $n \approx 3\text{-}6\times10^{19}/\text{cm}^3$ at very low temperatures. Because of the high doping concentration (larger than $\sim 2\times10^{19}/\text{cm}^3$), the Fermi energy ($E_F$) in Si:P lies in the conduction band[19], as in a metal (Supplementary Information Fig. S1a). Although the origin is not fully understood, it is well known that a small fraction of magnetic moments exists in metallic Si:P and persists even up to the degenerate doping concentration[7]. Thus, it is very likely that the local moments are entangled with the conduction electrons to form Kondo clouds that overlap with each other (see Fig. 1b), leading to a novel correlated ground state in the Si:P metal. In this work, tunnelling density-of-states (DOS) spectroscopy, together with bulk electrical transport measurements, allows us to explicitly identify the electronic and magnetic nature of competing ground states in Si:P metal and to construct the relevant phase diagram in the temperature ($T$)-magnetic field ($B$) plane.

We first measured the differential resistance $R_d$ of a bulk Si:P metal as a function of $T$. At $B = 0$, the $R_d$ shows the typical Fermi liquid $T^2$ temperature dependence at high temperatures (Fig. 2a). As $T$ decreases below approximately 2 K, a clear anomaly appears as a broadened step-like increase in $R_d$ with decreasing $T$. The differential resistance increases with $-\ln(T)$ behaviour (blue line), and then it seems to decrease slightly below ~ 1 K, which is very similar to the $T$ evolution of the resistivity of a Kondo lattice compound with disorder[20]. The observed $\ln T$ behaviour is typical of single-ion Kondo physics in metals with magnetic impurities[21]. A modest magnetic field switches the non-Fermi liquid metal state into a conventional disordered Fermi liquid (DFL), confirming that the observed resistivity anomaly below 2 K is associated with magnetic interactions.

In fact, the interplay between the competing interactions may switch the ground state of the Si:P metal. The competition between the ground states can be elucidated by observing the evolution of tunnelling DOS spectra near the $E_F$ of the Si metal at various $T$ and $B$ values. For this purpose, we fabricated tunnel junction devices consisting of the Si:P metal with silver (Ag) or aluminium (Al) electrodes separated by a thin $SiO_2$ tunnel barrier for the tunnelling DOS spectroscopy measurements (Figs. 2b, c). Because Ag behaves similar to a Fermi liquid with a flat DOS region in the vicinity of the $E_F$, the measured differential tunnelling conductance $G$ is directly proportional to the DOS of Si:P: $G(V) \propto DOS(E)$, where $E = E_F + eV$ ($V$ is the voltage across the tunnel barrier). Some of the basic results measured below 200 mK are presented in Figs. 2d, e, which display the measured $G$-$V$ characteristics at various temperatures at $B = 0$ and an intensity plot of $G$ in the $T$-$V$ plane, respectively. At $T = 18$ mK, a partial depletion of the tunnelling DOS near $E_F$, which may refered to as a pseudogap, is seen in the $G$-$V$ characteristics near $V = 0$, and anomalous peaks appear outside the pseudogap. As $T$ increases, the depleted DOS within the pseudogap is restored, and the peaks become closer in a nonlinear manner. Above ~160 mK, the U-shaped pseudogap with the side peaks changes into a $V^{1/2}$-type Altshuler-Aronov gap (also called a zero-bias anomaly: ZBA)[8], which is anticipated for metallic bulk Si with disorder and Coulomb correlations in the paramagnetic DFL phase.

Next, we present results that show how the anomalous DOS spectrum at the $E_F$ varies with $B$ at $T = 18$ mK. Figures 3a and b present examples of the measured $G$-$V$ characteristics at various $B$ values below 3000 Gauss (G) and an intensity plot of $G$ in the $B$-$V$ plane. As $B$ increases, the peaks become closer together, and the depth and width of the U-shaped pseudogap decrease. Most noticeably, the pseudogap and peaks smoothly change into the Altshuler-Aronov ZBA of the paramagnetic DFL phase, and afterwards, the $|V|^{1/2}$-dependent $G$-$V$ curve remains unchanged and independent of $B$. Interestingly, in the intermediate $B$ region between approximately 1000 and 2000 Gauss, $G(V)$ increases linearly as a function of $V$ up to the position of the considerably smaller peaks and then remains constant. Thus, this $|V|^{1}$-dependent $G$-$V$ curve has a V-groove shape, which is distinct from the ZBA. The inset in Fig. 3a shows a comparison of this V-groove pseudogap and the ZBA. The depth and width of the V-groove pseudogap continue to decrease with increasing $B$ and $T$ (Supplementary Information Fig. S2); however, its $B$-dependence is much weaker than that of the U-shaped pseudogap in the low $B$ region. We have confirmed that the observed anomalous DOS spectra are not hysteretic in $B$ by sweeping $B$ in both directions. Thus, the observed $B$-driven smooth phase transition, together with the electrical properties of a bulk Si:P metal, reveals that an exotic magnetically correlated state, which is not generally anticipated in a simple elemental semiconductor, seems to exist in the degenerately doped Si:P metal. $T$- and $B$-dependent characteristics of the pseudogap and anomalous peaks are presented in the Supplementary Information (Figs. S3 and S4).

Nonlinear behaviour in the $B$-dependent restoration of the DOS at $E_F$ ($G$ at $V = 0$) at various $T$ from 18 mK to 160 mK is observed (Supplementary Information Fig. S5). In the low $T$ region below approximately 100 mK, the DOS at $E_F$ is reinstated superlinearly with $B$ and then sublinearly up to the $B$ value, where the nonmagnetic $|V|^{1/2}$-dependent ZBA starts to appear in the DFL phase. There exists a clear inflection point in the derivative of the $G(V=0)$-$B$ curve, and we regard this point as the boundary between the superlinear (strongly $B$-dependent) and sublinear (weakly $B$-dependent) regions. As $T$ increases, the inflection $B$ point shifts to lower $B$ values, making a boundary in the $T$-$B$ plane (Supplementary Information Fig. S6).

The $T$-$B$ phase diagram of the Si:P metal shown in Fig. 4 summarizes our main experimental results. Two different phases in the intensity map of the DOS at the $E_F$ in the $T$-$B$ plane are clearly visible with an intervening region. The blue region corresponds to the magnetic metal phase where the U-shaped pseudogap and large anomalous DOS peaks outside the gap appear. The black diamonds show the inflection point in the derivative of the $G(V=0)$-$B$ curve, as a function of $T$. In this phase, the DOS spectrum at $E_F$ ($G$ at $V=0$) changes strongly as $B$ varies. The red stars represent the temperature, $T^*$, where the nonmagnetic $|V|^{1/2}$-type ZBA is detected first at different $B$ values, creating the most important characteristic boundary between the magnetically correlated metal phase and the nonmagnetic DFL phase. The magnetic metal phase transforms smoothly into the paramagnetic DFL phase (yellow region) through the intervening weakly magnetic metal phase (green region), where the V-groove DOS spectrum appears with considerably weaker DOS peaks. In this intervening phase, the depleted DOS at the $E_F$ is reinstated relatively weakly with increasing $B$. The boundaries consisting of the green circles indicate the temperature, $T_\Delta$, where the anomalous DOS peaks are first detected (Supplementary Information Fig. S3).

Observations of the abnormal electronic DOS spectra in the Si:P metal and their $B$-driven tuning are very surprising because neither the host Si nor the dopant P shows a magnetic order in the crystal structure, and degenerate Si:P is regarded as a metal (the $E_F$ is located within the conduction band). We rule out several possible mechanisms that could be responsible for our exotic DOS spectra. First, we rule out a Coulomb gap scenario because it cannot explain the magnetic behaviour. Second, we rule out spin glass-type magnetic gaps[22] because no noticeable differences are seen between the field-cooled and zero-field-cooled DOS spectra. Neither of these mechanisms can explain the existence of DOS peaks outside the gap or the observed non-Fermi-liquid behaviours in the bulk Si:P resistance. We also exclude the emergence of superconductivity. Unlike boron-doped degenerate Si[23], a clear signature of superconductivity (a collapse of the bulk resistance) is not detected down to 10 mK in the Si:P metal. We also eliminate the possibility of the Kondo lattice gap because the local moments in our sample are randomly distributed. Indeed, our DOS is highly symmetric, while the DOS of the Kondo lattice system has a characteristic asymmetry[24]. In fact, Si:P is not a heavy fermion system. Finally, the ground state with Ruderman-Kittel-Kasuya-Yosida (RKKY) interaction-induced magnetic order, for example, the random singlet state, is not relevant because it does not have any gap at the $E_F$[25]. Note that the local Kondo coupling dominates the RKKY interaction in our Si:P metal below 2 K, as we observe the Kondo effect.

All of the $T$- and $B$-dependent bulk transport, the DOS spectroscopy measurements and the existence of magnetic moments suggest that Kondo physics plays a key role in the formation of a correlated electron ground state in the Si:P metal. The underlying physics of the correlation behaviour is proposed here. That is, for a Si:P metal with a doping density of ~ $3\times10^{19}$/cm$^3$, more than a $10^{-5}$ fraction of the total impurities induce residual, unscreened localised moments below 2 K[7], and the mean interdistance between the moments is less than 1 μm, which may be comparable to or even smaller than the size of a Kondo cloud. Our proposed ground state is the condensation of overlapping Kondo clouds (Fig. 1b), and as a consequence of Kondo condensation, a fraction of itinerant electrons entangled with magnetic impurities are correlated to form a many-body ground state. This Kondo condensation model is fairly analogous to a Bose-Einstein condensate, and similar to BCS Cooper pairs, a new singlet ground state with a small BCS-like gap is formed in the DOS at the $E_F$. In fact, the shape and behaviour of the observed pseudogap in the Si:P metal is very similar to a BCS-like gap. In the lower doping regime where the metal-insulator transition (MIT) occurs, the newly discovered ground state has not been reported. This consequence is presumably because although the magnetic impurity density might be higher in that regime, the density of itinerant electrons to be captured by the magnetic impurities to form Kondo clouds is not enough. This effect is the reason why the observed Kondo phenomenon occurs in the degenerate metal regime, not in the impurity band regime.

Understanding many-impurity Kondo physics at a microscopic level is still challenging, especially for disordered solids. Nevertheless, the validity of the Kondo condensation scenario proposed here can be tested by simulating the electronic structure of the many-body singlet ground state. We reproduce the DOS with the correlation effect by considering the effect of condensation on the spectrum of itinerant fermions. For this purpose, we introduce a charge neutral scalar field $\varphi = <c_\uparrow f_\downarrow^+ - c_\downarrow f_\uparrow^+>$ where $c$ is the itinerant electron and $f^+$ is the impurity ion with net spin ½. We define Kondo condensation as the configuration where $\varphi$ is non-vanishing, and in this case, the value of $\varphi$ is our order parameter; its coupling to the fermion describes the effects of Kondo condensation on the spectrum of the fermion[26]. When $T$ or $B$ increases, the condensation is destroyed by thermal fluctuation or by Zeeman flips; thus, the pseudogap will disappear, and the system undergoes a transition from a gapped scalar ordered state, i.e., Kondo condensation, to the paramagnetic DFL state. It is a challenging problem to calculate the DOS spectra of this system in the presence of temperature and magnetic fields for a comparison with the observed data. The problem involves many sources of notorious difficulties in many-body theories, i.e., impurities and randomness in the presence of strong correlations. Proving the existence of nonvanishing condensation and calculating the spectrum starting from a local microscopic Hamiltonian is beyond our scope. It would be helpful if one can find an order parameter description of the system near the critical point and a mean field theory such that it works even for a strongly interacting system. One such theory is called holographic theory[9,10,11,27]. The validity of the theory relies on the universality of the system near the quantum critical point, and it encodes the quantum effects through classical gravity where the renormalization group[28] idea is implemented. Using this method, we calculated the spectral function in the presence of the scalar order. The outline is as follows: let $\psi(r,x)$ be the holographic fermion, which is dual to the actual fermion $\chi(x)$. Then, the Yukawa-type interaction, $\Phi(r,x)\bar\psi\psi(r,x)$, with canonical kinetic terms $\psi(r,x)$ and $\Phi(r,x)$ is used to calculate the spectral function[29]. More details of the calculation are given in the Supplementary Information. For the temperature evolution of the DOS spectra, we choose the parameters of the theory such that they can fit the medium temperature 100 mK data best.

Figure 5a reveals the modelled pseudogap in the DOS and two DOS peaks at opposite energies, calculated at $B = 0$. A gap is clearly visible in the DOS at $T = 18$ mK, whereas for a larger value of $T > 160$ mK, the pseudogap is closed, which is in qualitative agreement with the experimental $G$-$V$ curve at $B = 0$ and $T = 18$ mK (Fig. 2d). Because there is a nonzero DOS value at the $E_F$ ($V = 0$), the corresponding state is a metal. For larger magnetic fields, no strong pseudogap is present in the computed DOS (Fig. 5b). The calculated magnetic field-dependent DOS spectra are qualitatively similar to the experimental $G$-$V$ curves (Fig. 3a).

In conclusion, similarly shaped DOS spectra at the $E_F$ have been observed in the normal state of cuprate, pnictide and heavy fermion superconductors[15,16,30,31]. In fact, the $T$-$B$ phase diagram of our Si:P metal in Fig. 4 is also similar to that of the quantum materials listed above. This result is not a surprise because the observed pseudogap behaviours in these materials can be commonly explained by coherent electronic states with a correlation. Although a microscopic understanding of how the order parameter depends on external perturbations such as temperature, the magnetic field, pressure, and doping is still challenging, the observation of Kondo condensation and its phase transition in degenerately doped elemental semiconductor silicon will contribute to understanding other quantum materials such as Kondo lattices, spin glasses and high $T_C$ superconductors.

**Methods**

**Sample fabrication.** Ag-SiO$_2$-Si:P tunnel junction devices were fabricated on a [100]-oriented Si substrate, in which the phosphorus density ($n$) was 2-6×10$^{19}$/cm$^3$. The SiO$_2$ tunnel barrier was formed by thermal oxidation of the Si:P layer. The tunnel junction device was then completed by depositing a 200-nm-thick top Ag or Al electrode.

**Supplementary Information** is linked to the online version of the paper at www.nature.com/nature.

**Figures and Captions**

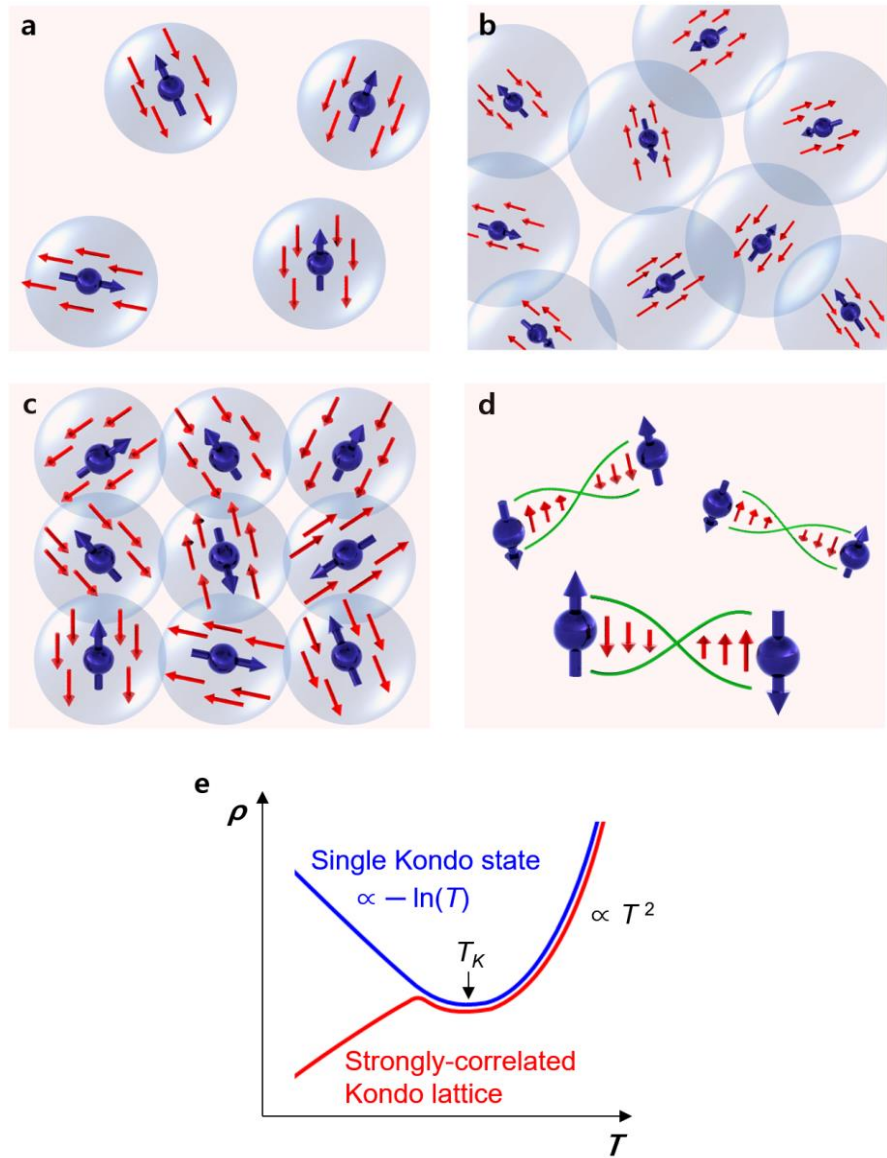

**Figure 1| Schematics of different singlet states in metals with magnetic impurities.** **a**, Unoverlapping Kondo singlet state where Kondo clouds are randomly distributed without interaction; this is practically the single-ion Kondo problem. **b**, Overlapping Kondo singlet state, i.e., Kondo condensation, where randomly distributed Kondo clouds overlap, interacting with each other and forming a correlated electron ground state. **c**, Kondo lattice where an electronic band of conduction electrons and a lattice of localised moments interact, forming a hybrid electronic structure. **d**, Random singlet state where two adjacent impurities interact via the RKKY interaction. The configuration of singlets in metals with magnetic moments is determined by many factors, such as the impurity density, randomness and complex

interactions between electrons and impurities. **e**, Schematic diagram illustrating the temperature-dependent resistivity $\rho(T)$ for two extreme cases: unoverlapping Kondo singlet state and correlated Kondo lattice. The single-impurity Kondo temperature $T_K$ is identified experimentally as the resistivity minimum. For a Kondo lattice with a dense periodic array of magnetic moments, overlapping Kondo screening clouds (Fig. 1c) lead to a heavy fermion state and a downturn in the resistivity.

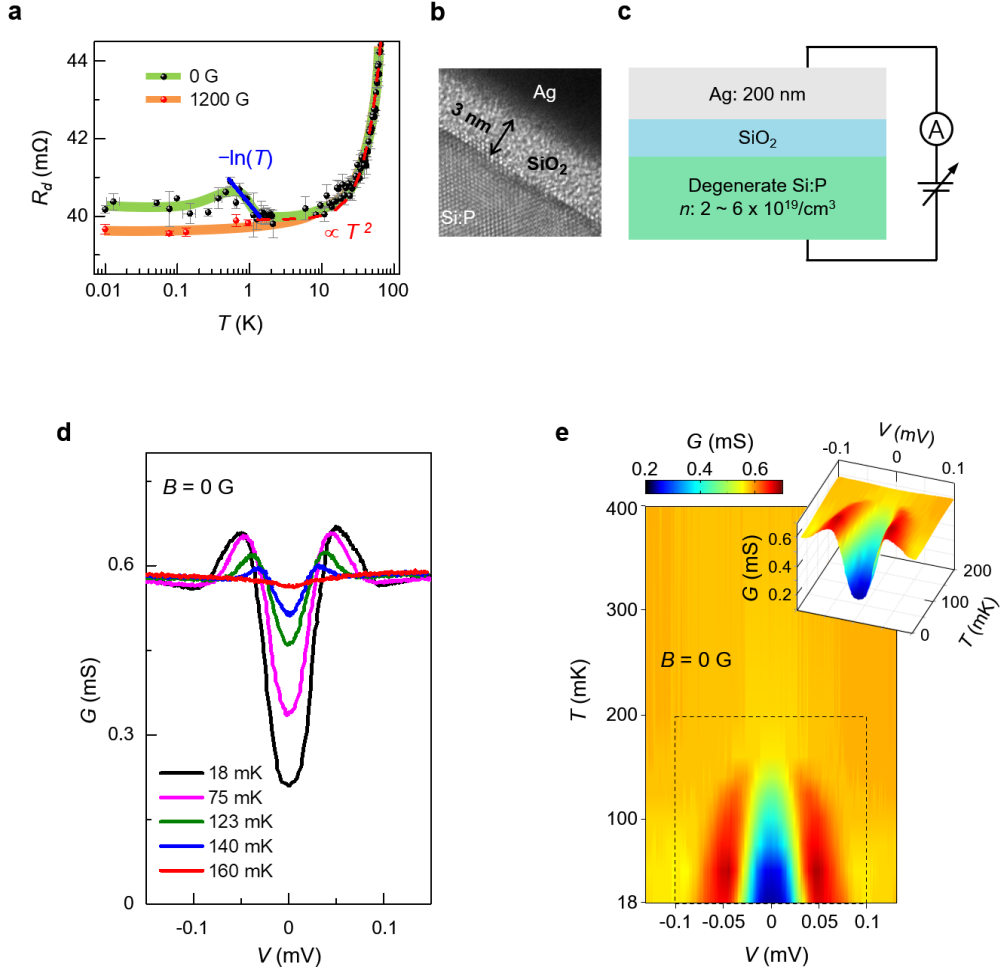

**Figure 2| Single-particle DOS spectra and the bulk resistance of a Si:P metal. a**, Temperature-dependent differential resistance $R_d$ of a bulk Si:P metal at $B = 0$ and 1200 Gauss, together with a fit according to the $T^2$-dependent Fermi liquid behaviour (red dotted line). **b**, Cross-sectional high-resolution transmission electron microscopy image of a Ag-SiO$_2$-Si:P tunnel junction device. **c**, Schematic of single-particle DOS measurements. **d**, The tunnelling conductance $G(V, B = 0)$ with decreasing temperature. Above ~ 160 mK, a $|V|^{1/2}$-type Altshuler-Aronov ZBA is observed in the DOS of Si:P near $E_F$. Upon cooling, a U-shaped pseudogap and DOS peaks are detected. **e**, Intensity map of the temperature-dependent $G(V, B = 0)$. The inset shows the three-dimensional enlarged view of the temperature-dependent $G(V, B = 0)$ curves in the low $T$ region.

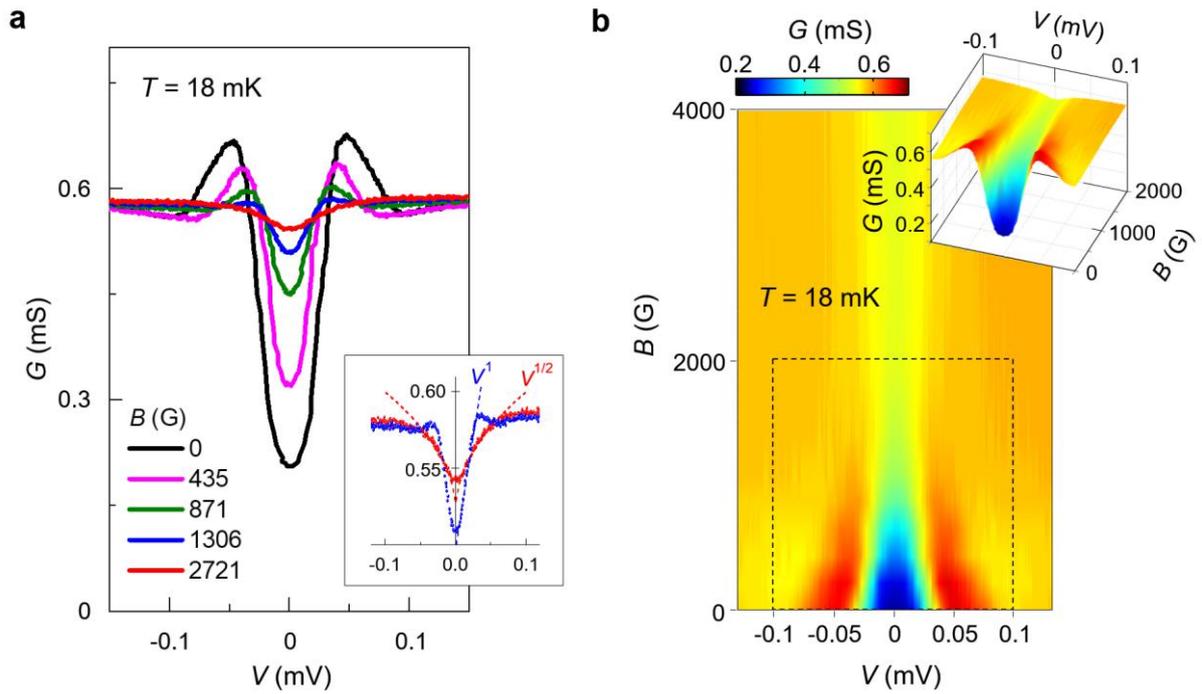

**Figure 3| Magnetic field-driven phase transition. a**, The tunnelling conductance $G(V, T = 18$ mK) with increasing magnetic field. The inset shows an enlarged view of the $G(V)$ curves measured at $B = 1306$ and $2721$ Gauss. The broken dotted lines are fitted to the curves using $|V|$ and $|V|^{1/2}$ relations. **b**. Intensity map of the $B$-dependent $G(V)$ at $T = 18$ mK. The inset shows the three-dimensional enlarged view of the $B$-dependent $G(V)$ curves in the low $B$ region.

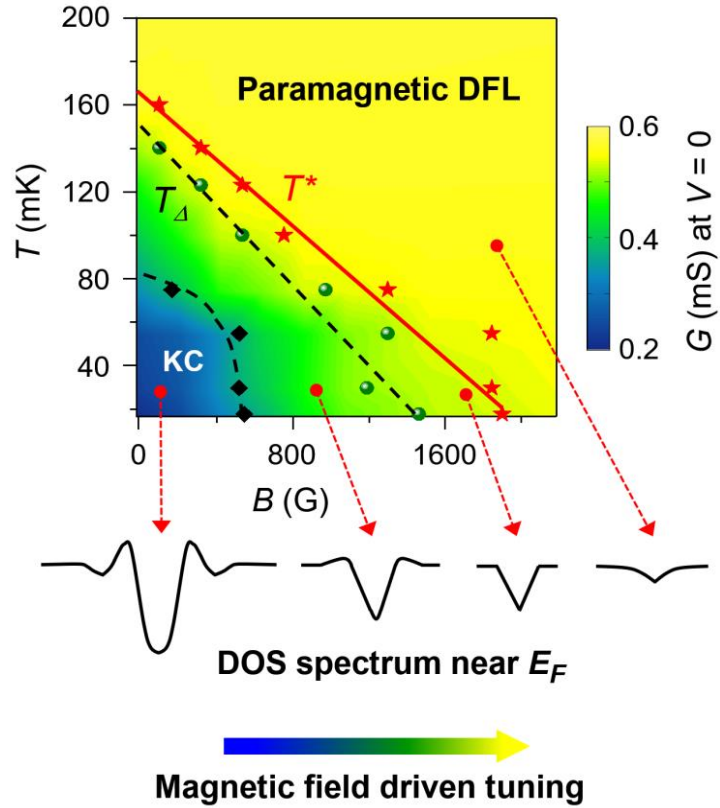

**Figure 4| Magnetic phase diagram of the Si:P metal.** Variation in the zero-bias conduction (namely, DOS at $E_F$ in Si:P) in the $T$-$B$ plane, showing the magnetic Kondo condensation (KC) phase (blue and green) and paramagnetic disordered Fermi liquid (DFL) phase (yellow). The magnetic transition line $T^*$ (red stars) and the boundary (black diamond dots) corresponding to the inflection points in the zero-bias $G(B)$ curves are plotted on the same phase diagram. See the text and Supplementary information Fig. S5. The green circles represent $T_\Delta$ (Supplementary information Fig. S3a). The applied magnetic field controls the magnetic correlations in the Si:P metal. Characteristic DOS spectra at $E_F$ of Si:P in each phase region are shown.

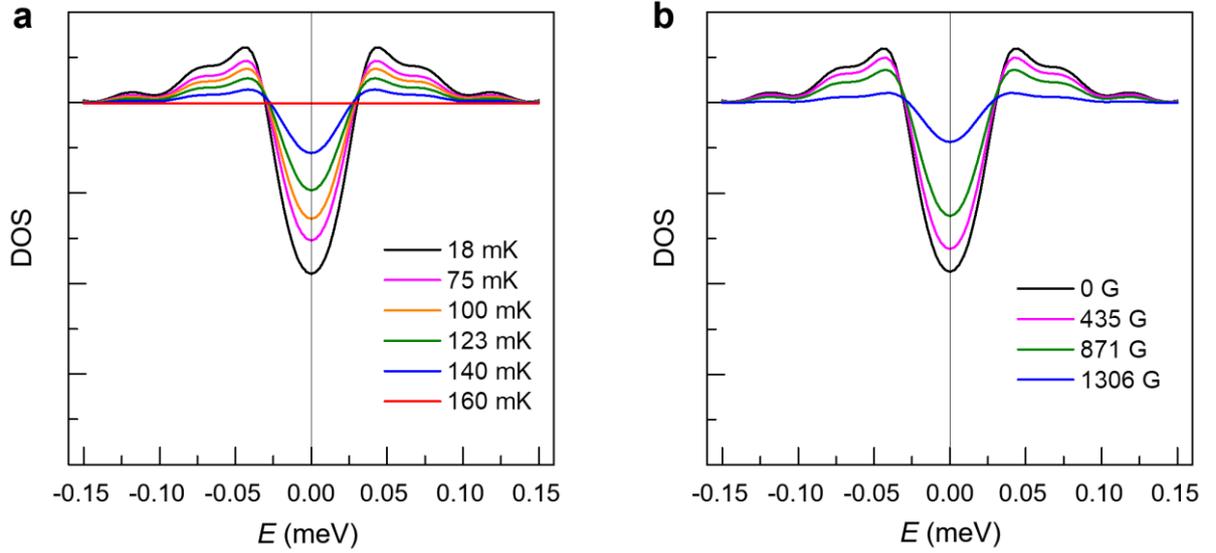

**Figure 5| Theoretically calculated DOS spectra. a,** Calculated DOS as a function of $T$ at $B = 0$. **b,** Calculated DOS as a function of $B$ at $T = 18$ mK. Here, $E = 0$ corresponds to the Fermi energy. For the data fitting at different temperatures, we modelled the temperature dependence of the condensation parameter by $M(T) = M_0(1-T/T^*)^{1/2}$, which is universally true for any mean field theory including holographic theory. We choose $M_0$ such that the theory fits the data at a particular temperature best, which we take $T = 100$ mK. Then, the same parameters were used to see how the theory can fit the data for other temperatures. Similarly, for the magnetic field evolution, we modelled the condensation's magnetic field dependence as $M(T) = M_0(1-T/T^*-B/B_c)^{1/2}$, where $T^*$ is the critical temperature at $B = 0$ and similarly $B_c$ is the critical field at $T = 0$. For calculations, we take $T^* = 147$ mK, $B_c = 1600$ G and $M_0 = 5.44$. The magnetic phase boundary in the $T$-$B$ plane (Fig. 4) is simply given by the $M = 0$ curve: $T/T^* + B/B_c = 1$.

# Supplementary Information

# Observation of Kondo condensation in a degenerately doped silicon metal

## S1. Sample preparation

*Material and Device*. Ag-SiO$_2$-Si:P and Al-SiO$_2$-Si:P tunnel junction devices were fabricated on a [100]-oriented degenerate Si:P substrate. The SiO$_2$ tunnel barrier was formed by thermal oxidation of the Si:P layer. The tunnel junction device was then completed by depositing 200-nm-thick top Ag (or Al) electrodes. The diameter of the Ag (or Al) electrodes was in the range of 200 ~ 500 μm. The measured resistivity of Si:P ranges from 0.001 - 0.003 Ωcm at 300 K, which corresponds to a doping concentration (*n*) of 2-6 × 10$^{19}$/cm$^3$ (Ref. [S1]). This doping level is consistent with the P concentration obtained from the secondary ion mass spectrometry (SIMS) depth profile results for Si:P (3-6 × 10$^{19}$/cm$^3$). The measured carrier concentration of Si:P at 4 K was also between 3-6 × 10$^{19}$/cm$^3$. All these data confirm that Si:P is degenerate. No magnetic impurity atoms were observed in the SIMS analysis. Transmission electron microscopy measurements reveal that Si:P is crystalline.

*Bulk resistance measurements*. The temperature- and magnetic field-dependent differential resistance $R_d$ of Si:P (typical dimensions: 6 mm (*l*) × 0.4 mm (*w*) × 0.6 mm (*h*)) was measured using a four-point lock-in technique with a small excitation current of 1 μA (frequency = 79 Hz). Al electrodes (200-nm-thich) were used for the bulk resistance measurement.

## S2. Comparison of the tunnelling DOS spectra of Si:B near the MIT region and degenerately doped Si:P metal

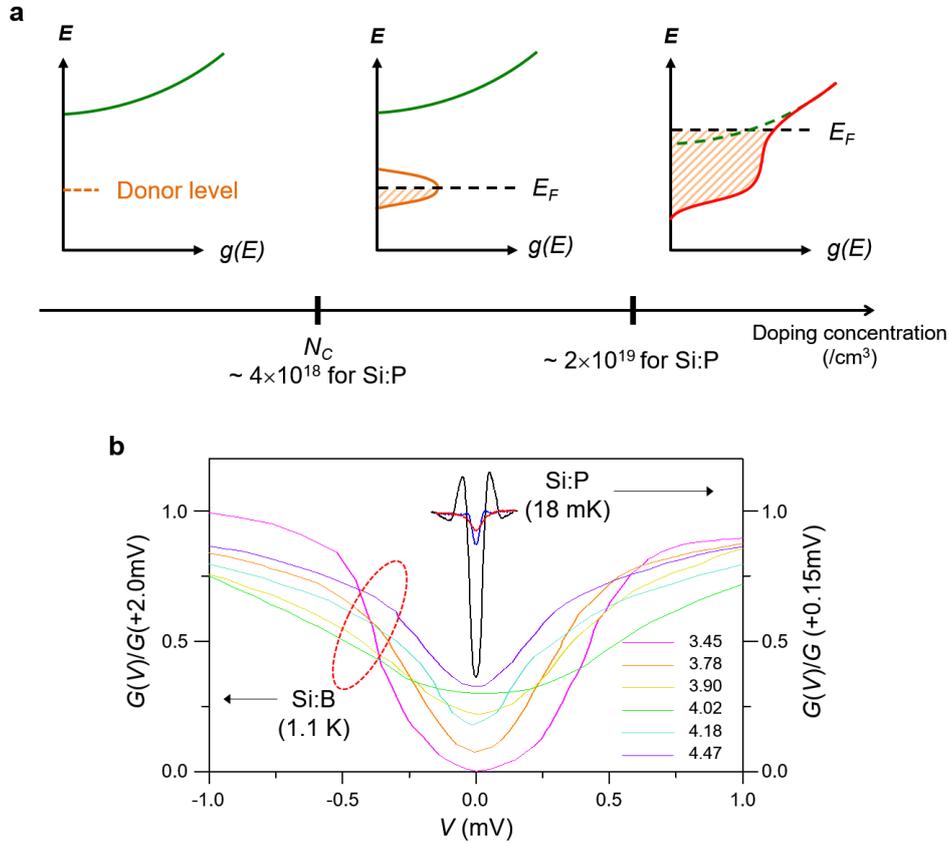

**Figure S1. a**, Energy vs. density of states (DOS) in Si doped with different doping densities ($N$). $N_C$ is the critical doping concentration for the metal insulator transition (MIT). **b**, Tunnelling DOS spectroscopy in Si:B near the MIT region (doping concentration $N$ ranges between 3.45 and 4.47×$10^{18}$/cm$^3$) [J. G. Massey & M. Lee, *Phys. Rev. Lett.* **77**, 3399 (1996).] and in our degenerately-doped Si:P metal ($N > 2 \times 10^{19}$/cm$^3$). The tunnelling DOS spectra were measured at $B = 0$. The shape of the DOS spectra in the Si:P metal differs from that of the spectra in Si:B, suggesting that the fundamental physics for the intriguing DOS in the Si:P metal is completely different. The measured DOS spectra in the MIT region are nonmagnetic, whereas the DOS spectra in the Si:P metal are magnetic.

## S3. Evolution of the V-groove pseudogap and zero-bias anomaly (ZBA)

Figure S2 presents the evolution of the cusp-like Altshuler-Aronov ZBA and the V-groove spectrum along different paths in the $T$-$B$ phase diagram (Fig. 4). As $T$ increases to 160 mK (orange arrow in the phase diagram), the ZBA spectrum does not change in shape, confirming its nonmagnetic origin. On the other hand, as $B$ increases to 1524 Gauss (green arrow in the phase diagram), the V-groove pseudogap smoothly changes into the ZBA spectrum of the DFL phase.

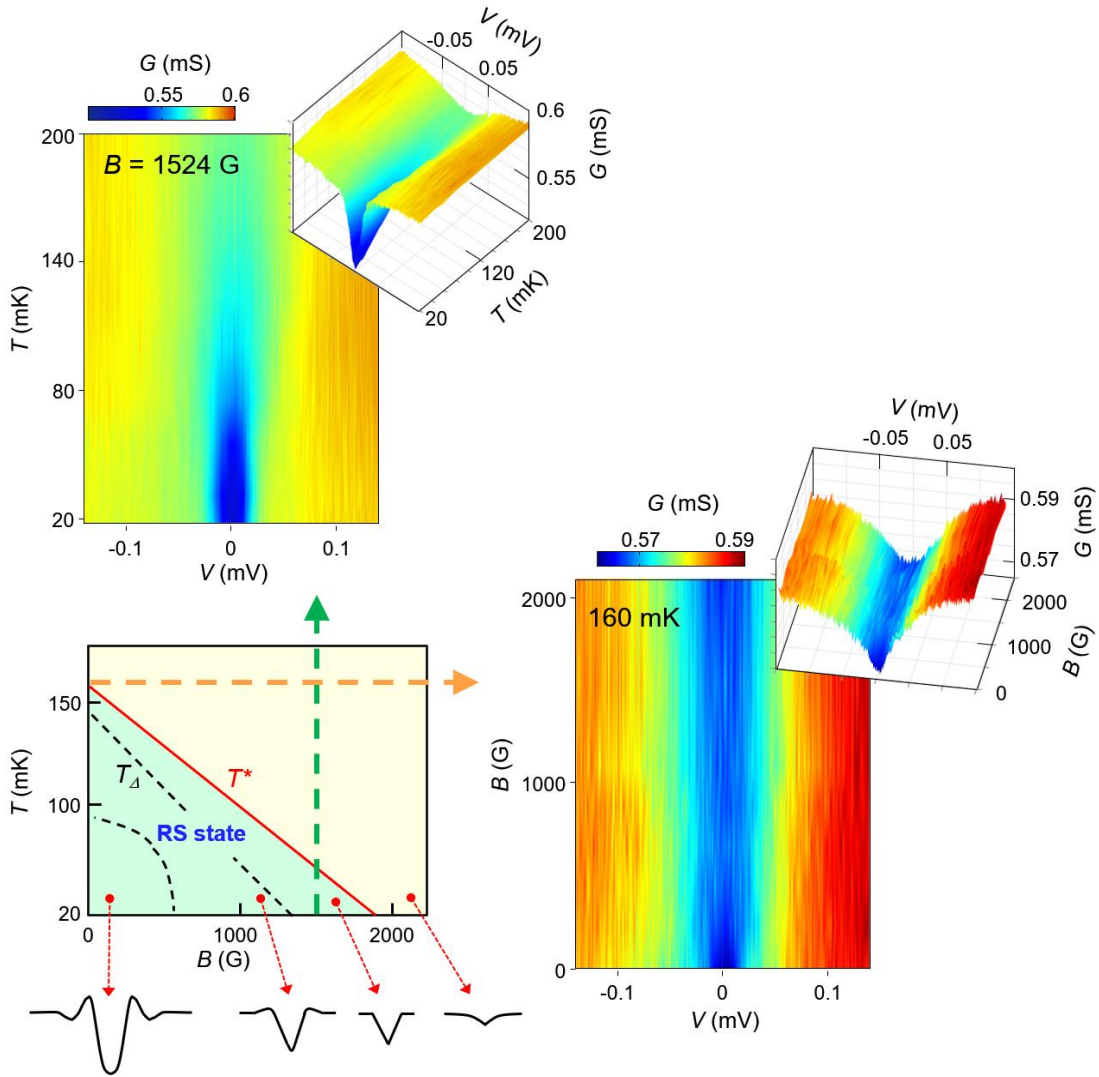

**Figure S2.** Intensity maps of the $G(V)$ curves showing the evolution of the ZBA spectrum and the V-groove spectrum along the different path in the $T$-$B$ phase diagram.

## S4. Characteristics of the pseudogap and anomalous DOS peaks

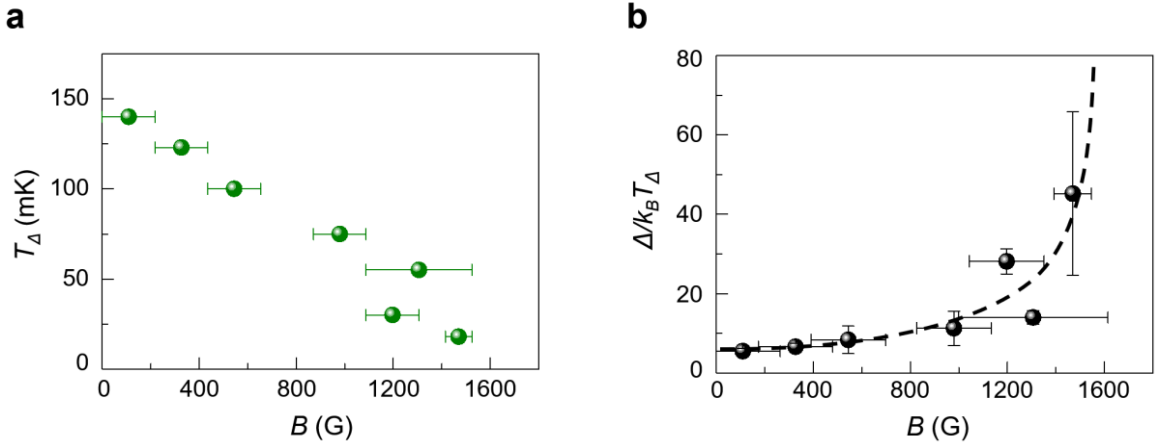

**Figure S3. a**, Maximum temperature, $T_\Delta$ where the anomalous DOS peaks are detected, as a function of $B$. **b**, $\Delta/k_B T_\Delta$ versus $B$. $k_B$ is Boltzmann's constant.

Let $\Delta$ be the pseudogap in units of energy, defined as the separation of the two measured peaks and $T_\Delta$ be the maximum temperature where the anomalous DOS peaks are detected. Then $T_\Delta$ decreases with increasing $B$ (Fig. S3a). As $T$ and $B$ increase, the peaks shift systematically to lower bias voltages, and the peak height decreases similarly. The $T$- and $B$-dependence of the pseudogap size and the peak height ($\Lambda$) are presented in the Supplementary Information Fig. S4. Similarities between $\Delta(T, B)$ and $\Lambda(T, B)$ are observed, suggesting that they have the same origin. The value of $\Delta/k_B T_\Delta$ increases monotonically with increasing $B$ in the low $B$ region below approximately 700 Gauss before increasing dramatically at higher $B$ (Fig. S3b). The steep increase in $\Delta/k_B T_\Delta$ in Fig. S3b is consistent with $T_\Delta \to 0$ K at $B \sim 1600$ Gauss which is shown in Fig. S3a.

## S5. Temperature and magnetic field dependences of the pseudogap depth and anomalous peak height

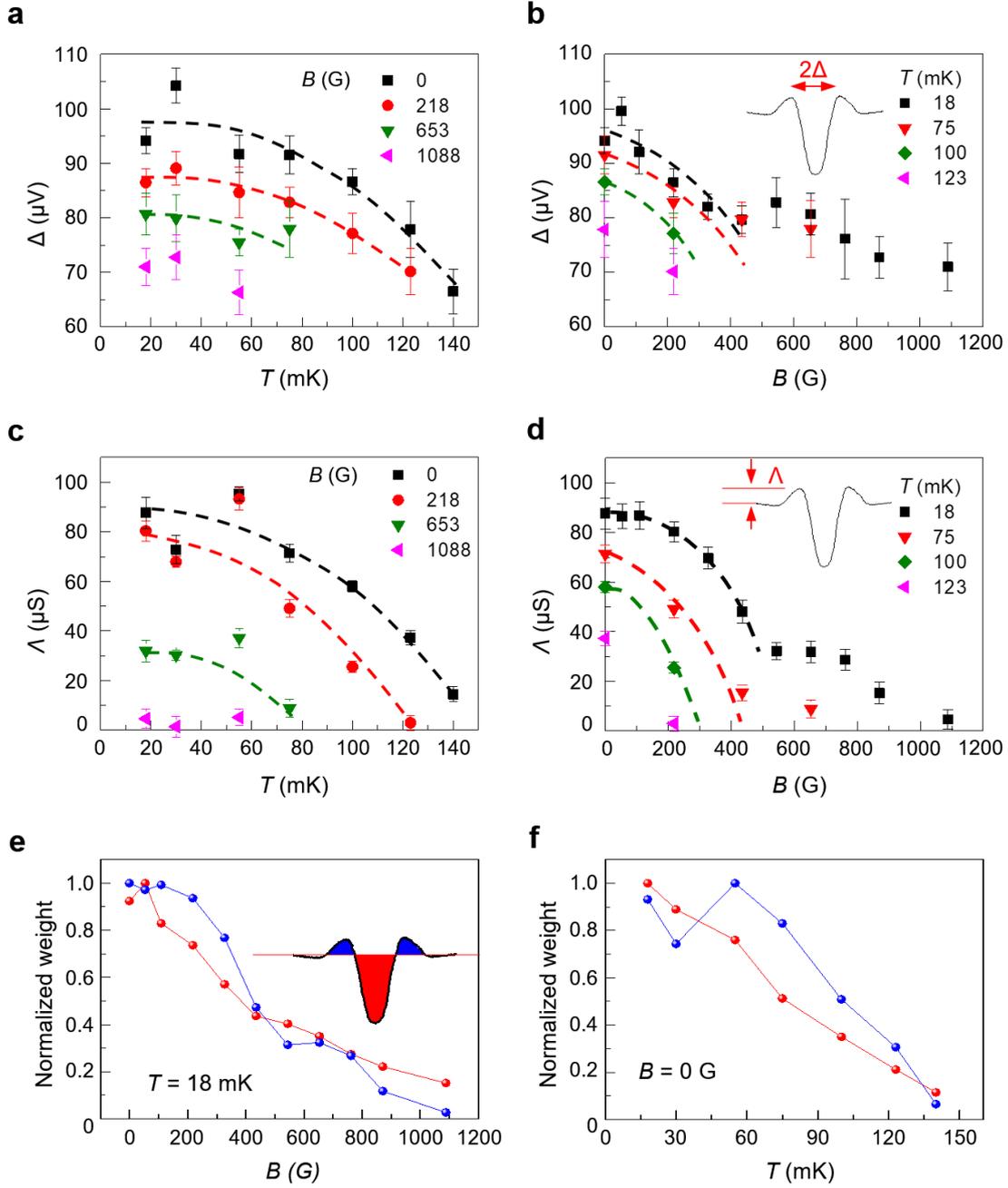

**Figure S4. a**,**b**, Pseudogap size $\Delta$ as a function of $T$ and $B$. **c**,**d**, Peak height $\Lambda$ as a function of $T$ and $B$. The peak height is measured from the flat baseline at a higher bias voltage. The dashed lines are a guide to the eye. **e**,**f**, Temperature and magnetic field dependences of integrated pseudogap depth and peak height, respectively. The correlation between the pseudogap at $E_F$ and the anomalous peaks outside the gap is further corroborated by the temperature and magnetic field dependences of the integrated pseudogap depth (red region) and anomalous peak height (blue regions). The integrated pseudogap depth and peak amplitude decrease similarly with increasing temperature and magnetic field. The voltage range of integration is taken from crossings between the DOS spectra and the extended flat DOS line of the high voltage region (red line in the inset in Fig. S4e).

**S6. Intensity plot of the *B*-dependent values of *G*(*V*, *B*) in the *G-B* plane**

To further demonstrate the smooth magnetic phase transition from the magnetic pseudogap phase to the paramagnetic DFL phase, another intensity plot for the zero-bias differential conductance, *G*(*V*,*B*) curves at 18 mK is presented in Fig. S5 for each value of *B*. The upper and lower boundaries of this plot correspond to the DOS maximum (anomalous DOS peaks) and DOS minimum (pseudogap) at the $E_F$, respectively. As the magnetic field *B* increases, the magnetic phase with the U-shaped pseudogap transforms smoothly into the paramagnetic DFL phase with the Altshuler-Aronov ZBA. Beyond approximately 1959 Gauss, the paramagnetic ZBA appears independent of *B*.

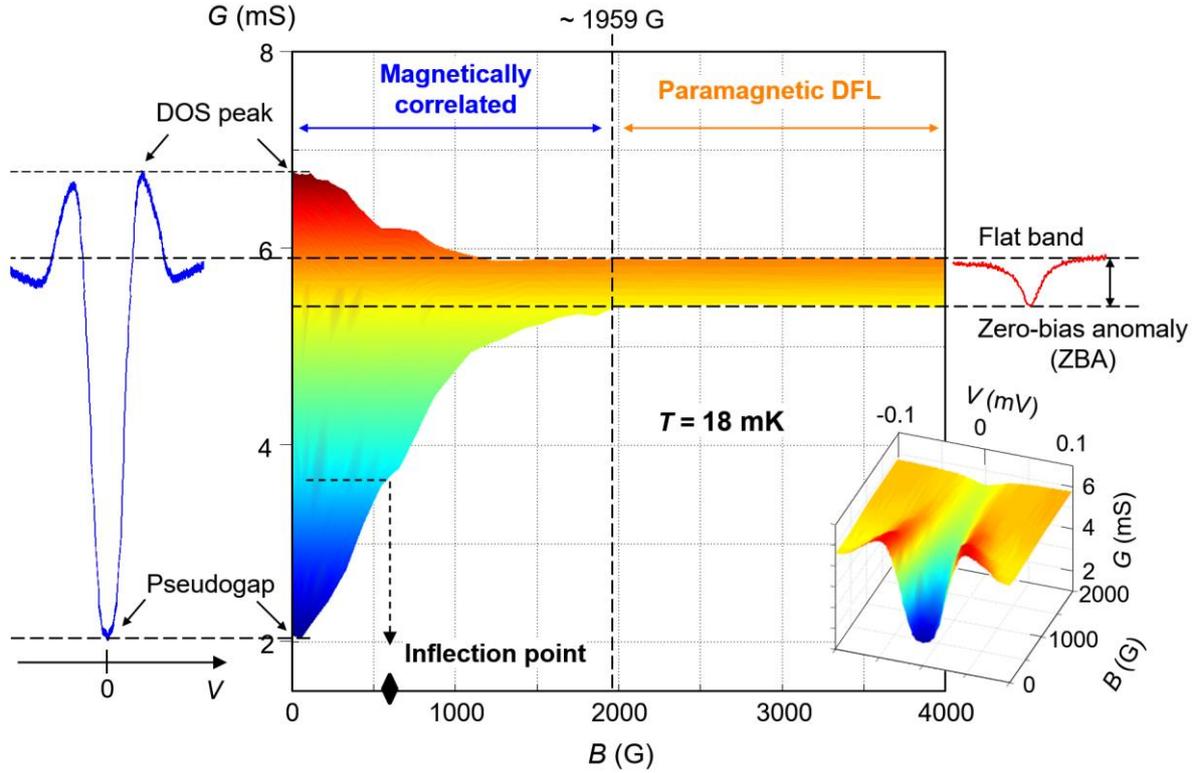

**Figure S5.** Intensity plot of $G(V, B)$ for each $B$ at 18 mK.

### S7. *B*-dependence of the zero-bias *G* value

The nonlinear behaviour in the *B*-dependent restoration of the DOS at the $E_F$ ($G$ at $V = 0$) at various *T* values from 18 mK to 160 mK is plotted in Fig. S6a. In the low *T* region below approximately 100 mK, the DOS at $E_F$ is reinstated superlinearly with *B*, and then sublinearly up to the *B* value where the nonmagnetic $|V|^{1/2}$-dependent ZBA starts to appear in the DFL phase. As shown in the inset, there exists a clear inflection point (marked with vertical arrows) in the derivative of the $G(V=0)$-*B* curve, and we regard this point as the boundary between the superlinear (strongly *B*-dependent) and sublinear (weakly *B*-dependent) regions. Figure S6b shows the inflection point as a function of *T* (black diamond). As *T* increases, the inflection point *B* shifts to lower *B* values, making a boundary in the *T*-*B* plane. The red stars represent the temperature, $T^*$, where the nonmagnetic $|V|^{1/2}$-type ZBA is detected first at different *B* values, creating the most important characteristic boundary between the magnetically correlated metal phase and the paramagnetic DFL phase.

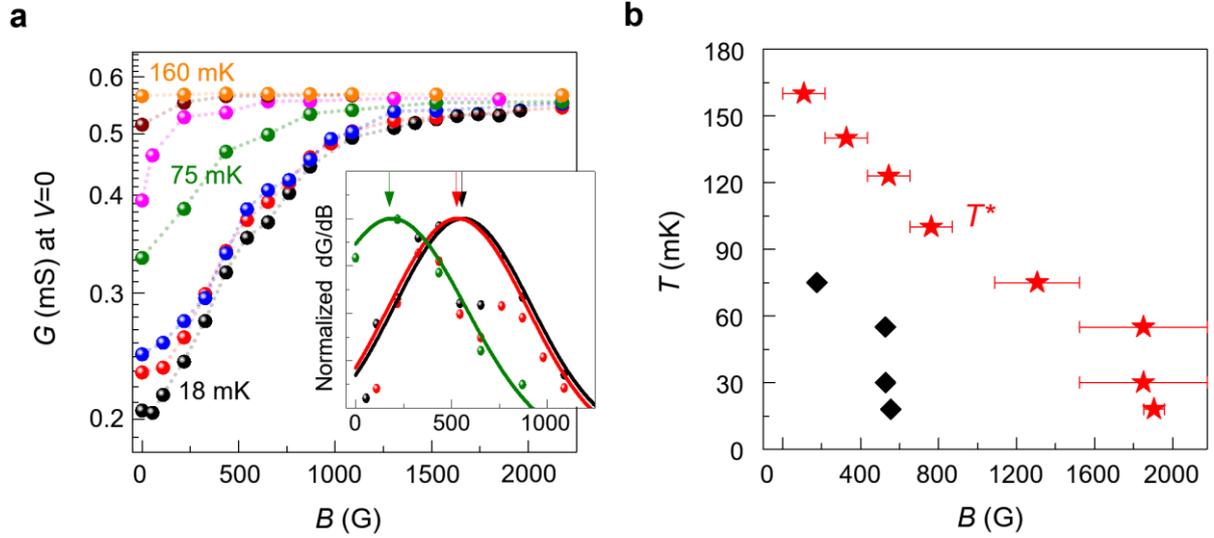

**Figure S6. a**, $B$-dependence of the zero-bias $G$ value (DOS at $E_F$ in Si:P) at various $T$ values from 18 mK to 160 mK. The inset shows the derivative of the zero-bias $G(B)$ curves (solid dots) with the best Gaussian fits to the curves (solid lines). The arrows indicate the inflection points of the derivative curves. **b**, The inflection $B$-point (black diamond) of the zero-bias $G(B)$ curves in a and the magnetic pseudogap transition line $T^*$ (red star) are plotted in the $T$-$B$ plane.

## S8. Holographic calculation of the density of states

### S8.1. Strong correlation and holographic field theory

There are two ways for a condensed matter system to be strongly correlated. One is to have a small Fermi surface (FS); that is the Coulomb interaction in a metal is small only because of charge screening, which is efficient in the presence of a large FS. Therefore, if the FS is small, screening cannot be effective in diminishing the interaction. Extremely clean monolayer graphene[S2,S3] and the surface of a topological insulator[S4-S6] provide examples of systems showing anomalous transport that can be explained holographically[S7-S9]. In fact, any Dirac material is strongly correlated as far as its FS is near the tip of the cone of the electronic band structure.

Another way for a system to be strongly correlated is to have slowly moving electrons. A well-known example in this class is the transition metal oxides such as cuprates, where the outermost electrons of the 4s-electrons are taken by the oxygen atoms, thereby suppressing the hopping of the electrons in 3d shells because the electrons have to tunnel a larger distance. The second in this class is the disorder induced strong correlation, where electrons are semi-localised by the Kondo physics. Our system is a typical example of this class.

Once particles strongly interact, all the particles in the entire system become entangled, and a single-particle picture cannot be used to understand the low-energy properties of the system. In such a case, reducing the degrees of freedom involved in determining the low-energy properties is the key issue of the interacting quantum many-body systems. One such idea is to look at the quantum critical point (QCP), where there is no scale so that no lattice specific information can be there. Detailed knowledge of the system cannot be retained there, and almost all information is lost except for a few critical exponents. Such information loss leads to universality, which is very similar to the black hole (BH) case: the no-hair theorem states that apart from the conserved total electrical charge, a BH does not have many characteristics. Such an analogy between a QCP and a BH is the guiding principle of the holographic approach for strongly correlated systems in condensed matter physics[S10,S11]: if a QCP is characterized by the parameters $z$ and $\theta$, which are associated with the dispersion relation $\omega \sim k^z$ and the entropy density $s \sim T^{(d-\theta)/z}$, then there exists a metric with the same scaling symmetry:

$$ds^2 = r^{-\theta}\left(-r^{2z}f(r)dt^2 + \frac{dr^2}{r^2 f(r)} + r^2(dx^2 + dy^2)\right), \quad \text{(C. 1)}$$

with $(t, r, x) \to (\lambda^z t, \lambda^{-1} r, \lambda x)$ and the method developed for the exact Anti-de Sitter/Conformal Field Theory (AdS/CFT)[S12-S14] is assumed to hold for this case. Here we consider the $z = 1$ case for simplicity.

## S8.2. Spectral function and density of state in holographic theory

The action for the fermion and the real scalar in d+1 dimensional spacetime is given by[S15]

$$S_D = \int d^{d+1}x\sqrt{-g}\bar{\psi}(\Gamma^M D_M - m - \Phi)\psi + \int d^{d+1}x\sqrt{-g}(|\partial_\mu \Phi|^2 - m^2\Phi^2) \quad (C.2)$$

where $m$ is the bulk mass of the scalar $\Phi$, $g$ is the determinant of the metric tensor $g_{MN}$, and $\Gamma^M = E^M{}_a \Gamma^a$ and $\Gamma^a$ are the gamma matrices. The vielbein $E^M{}_a$ is defined by $E^M{}_a E^N{}_b \eta^{ab} = g_{MN}$. The subscript $D$ denotes the Dirac fermion and the covariant derivative is given by

$$D_M = \partial_M + \frac{1}{4}\omega_{abM}\Gamma^{ab} - iqA_M, \quad (C.3)$$

where $\omega_{abM}$ is the spin connection and $A_M$ is the vector potential, which is dual to the electron num.

For fermions, the equation of motion is a first-order differential equation, and we cannot fix the values of all the components at the boundary, which makes it necessary to introduce the 'Gibbons-Hawking term' $S_{bd}$ to guarantee the equation of motion which can be written as

$$-iS_{bd} = \pm\frac{1}{2}\int d^d x\sqrt{h}\bar{\psi}\psi = \pm\frac{1}{2}\int d^d x\sqrt{h}(\bar{\psi}_-\psi_+ + \bar{\psi}_+\psi_-), \quad (C.4)$$

where $h = -gg^{rr}$ and $\psi_\pm$ are the spin-up and spin-down components of the bulk spinors, respectively. The overall sign ± is chosen such that when we fix the value of $\psi_+$ at the boundary, $\delta S_{bd}$ cancels the terms including $\delta\psi_-$ that arises from the total derivative of $\delta S_D$. A similar approach is true when we fix $\psi_-$. The former defines the standard quantisation, and the latter performs the alternative quantization. The gravitational solution is Reisner-Nordstrom BH in an asymptotic $AdS_4$ spacetime, which may be considered a deformation of the metric for the case of $z = 1$ and $\theta = 0$ by a possible presence of the electrical charge.

$$ds^2 = -r^2 f(r)dt^2 + \frac{1}{r^2 f(r)}dr^2 + r^2 d\vec{x}^2 \quad (C.5)$$

$$f(r) = 1 - \frac{r_0^3}{r^3} - \frac{r_0\mu^2}{r^3} + \frac{r_0^2\mu^2}{r^4} \,. \tag{C.6}$$

For the *RN-AdS*$_4$ BH, the horizon $r_0$ is defined as

$$r_0 = \frac{1}{3}(2\pi T_B + \sqrt{4\pi^2 T_B^2 + 3\mu^2}) \tag{C.7}$$

where $T_B$ is the BH temperature chosen to be $T_B = 0.78$ and $\mu = 2$ for our system. The asymptotic behaviour of $\Phi$ in the metric is

$$\Phi = \frac{\Phi^{(0)}}{r} + \frac{\Phi^{(1)}}{r^2} + \cdots \tag{C.8}$$

Here $\Phi$ is a real scalar field that does not couple with a gauge field $A_\mu$. For $\Phi^{(0)} = 0$, $\Phi^{(1)}$ has the form of

$$\Phi^{(1)} = M_0\sqrt{1 - T/T^*} \tag{C.9}$$

where $T^*$ is the critical temperature. We considered $T$ to be the temperature of the laboratory, and the condensation of $\Phi^{(1)}$ should be zero if $T > T^*$.

Following ref. [S15], we now introduce $\phi_\pm$ by

$$\psi_\pm = (-gg^{rr})^{-\frac{1}{4}}\phi_\pm, \quad \phi_\pm = \begin{pmatrix} y_\pm \\ z_\pm \end{pmatrix}. \tag{C.10}$$

Then the equations of motion, after Fourier transformation, can be written as[S15] follows:

$$\sqrt{\frac{g_{xx}}{g_{rr}}} z'_+(r) - \left(m + \frac{M}{r^2}\right)\sqrt{g_{xx}} z_+(r) + i[u(r) + k_x]y_-(r) - ik_y z_-(r) = 0$$

$$\sqrt{\frac{g_{xx}}{g_{rr}}} y'_-(r) + \left(m + \frac{M}{r^2}\right)\sqrt{g_{xx}} y_-(r) + i[u(r) - k_x]z_+(r) - ik_y y_+(r) = 0$$

$$\sqrt{\frac{g_{xx}}{g_{rr}}} y'_+(r) - \left(m + \frac{M}{r^2}\right)\sqrt{g_{xx}} y_+(r) - i[u(r) - k_x]z_-(r) + ik_y y_-(r) = 0$$

$$\sqrt{\frac{g_{xx}}{g_{rr}}} z'_-(r) + \left(m + \frac{M}{r^2}\right)\sqrt{g_{xx}} z_-(r) - i[u(r) + k_x]y_+(r) + ik_y z_+(r) = 0$$

$$\text{(C.11)}$$

where $u(r) = \sqrt{\frac{g_{xx}}{-g_{tt}}}(\omega + qA_t(r))$. Here, $k_x$ is the momentum along the $x$ direction. Near the boundary ($r \to \infty$), the geometry becomes $AdS_4$ and the equations of motion have the following analytical solution

$$z_+ = A_1 r^m + B_1 r^{-m-1} + \cdots, \quad y_- = C_1 r^{m-1} + D_1 r^{-m} + \cdots, \quad \text{(C.12)}$$

$$y_+ = A_2 r^m + B_2 r^{-m-1} + \cdots, \quad z_- = C_2 r^{m-1} + D_2 r^{-m} + \cdots. \quad \text{(C.13)}$$

Here, we made an abbreviation for mL with m, which should be restored at the end of the calculations (see S7.3)

The boundary term in Eq. (C. 4) becomes

$$-iS_{bd} = z_+^* y_- - y_+^* z_- = (A_1^* D_1 - A_2^* D_2) + \sum_\pm E_\pm r^{\pm 2m-1} + E_2 r^{-2}, \quad \text{(C.14)}$$

using the asymptotic behaviour of wave functions $\chi_i$. Here, $E_\pm$ and $E_2$ are functions of A, B, C, and D. Then, the retarded Green's function in the standard quantisation condition is given by

$$\tilde{\mathcal{G}} = \text{diag}\left(i\frac{A_1}{D_1}, -i\frac{A_2}{D_2}\right) \equiv \text{diag}(\tilde{G}_+^R, \tilde{G}_-^R) = -\text{diag}\left(\frac{1}{G_+^R}, \frac{1}{G_-^R}\right), \quad m > 0. \quad \text{(C.15)}$$

Since $G_R$ for the $m < 0$ case can also be obtained by the substitution $G_R \to -1/G_R$, $\tilde{G}_R$, Green's function for the alternative quantisation for $m > 0$ is the same as that for $-m$ in the standard quantization:

$$\tilde{G}_\pm^R(\omega, k; m) = -\frac{1}{G_\pm^R(\omega, k; m)} = G_\mp^R(\omega, k; -m). \quad \text{(C.16)}$$

The spectral function is defined as the imaginary part of Green's function. There are two of them $\Im[G_+^R]$ and $\Im[G_-^R]$ and we can define the spectral function for each of them:

$$A_\pm(\omega, k) = \Im[G_\pm^R(\omega, k)]. \quad \text{(C.17)}$$

There is an issue regarding the finiteness of the spectrum. It was pointed out that the high-frequency behaviour of the spectral function diverges similar to $\omega^{2m}$ so that the sum of the degrees of freedom over the frequency is infinite if *m* is positive. Therefore, we need to take the negative bulk mass only in the standard quantisation. For ease of discussion, we want to maintain the positivity of the mass, which can be done simply by going to the alternative quantisation. Even in this case, the spectral function is $\sim \omega^{-2m}$, which does not decay fast enough to guarantee finite integration. The sum rule can still be an issue[S16,S17], and we do not treat this problem here. To summarize, we work in alternative quantisation with a positive mass, and we treat the fermion as a probe and do not consider its back-action.

**S8.3. Conversion**

To compare the experimental data, we needed to use dimensionful parameters, while our formula was written in terms of dimensionless variables in the natural unit where $\hbar = k_B = v_F = 1$. The prescription to restore the units is

$$T \to \frac{k_B T}{\hbar v_F} L = \frac{\hat{T}}{Kelvin} \frac{L}{2.3 \times 10^6 nm} \tag{C.18}$$

$$B \to \frac{e}{\hbar} B L^2 = \frac{\hat{B}}{Tesla} \frac{L^2}{(25.7 nm)^2} \tag{C.19}$$

$$M_0 \to \frac{M_0}{(\hbar v_F)^2} L^2 = \frac{9 v_0^2}{4 v_F^2} \frac{L^2}{(nm)^2} \frac{M}{(eV)^2} \tag{C.20}$$

where *L* is an arbitrary length scale that cancels everywhere in the formula and $v_0$ = (the speed of light)/300. Let us finish with a concluding remark: The P atoms that possess local moments are just a fraction of the total number of phosphorous atoms. Therefore, we should think of our system as consisting of two fluids: one system includes the electrons involved in the RKKY interaction with the localised moments and the other is the electron system that does not interact with the localised moments. The BH has a temperature that is the temperature of the whole electron system, which is denoted as $T_B$. However, the condensation temperature of the paired localised moments does not necessarily coincide with $T_B$. In fact, it is assumed that $T_B$ does not change over the scale where the condensation temperature changes from 18 mK to 200 mK because the BH temperature, which describes the equilibrium temperature, should change more slowly than the transient temperature describing the quantum coherent system. In fact, it does not make sense to define the equilibrium temperature of a quantum coherent system because the thermal system is a mixed state, while the coherent state is a pure state. Therefore, the temperature of the coherent system is just a local transient temperature.